\documentclass[a4paper]{report}
\usepackage[utf8]{inputenc}
\usepackage[T1]{fontenc}
\usepackage{RJournal}
\usepackage{booktabs}

\begin{document}

\sectionhead{Contributed research article}
\volume{XX}
\volnumber{YY}
\year{20ZZ}
\month{AAAA}

\begin{article}
\title{\texttt{ceylon}: An R package for plotting the maps of Sri Lanka}
\author{by Thiyanga S. Talagala, Department of Statistics, Faculty of Applied Sciences, \break University of Sri Jayewardenepura}

\maketitle

\abstract{%
The rapid evolution in the fields of computer science, data science, and
artificial intelligence has significantly transformed the utilisation of
data for decision-making. Data visualisation plays a critical role in
any work that involves data. Visualising data on maps is frequently
encountered in many fields. Visualising data on maps not only transforms
raw data into visually comprehensible representations but also converts
complex spatial information into simple, understandable form. Locating
the data files necessary for map creation can be a challenging task.
Establishing a centralised repository can alleviate the challenging task
of finding shape files, allowing users to efficiently discover
geographic data. The ceylon R package is designed to make simple feature
data related to Sri Lanka's administrative boundaries and rivers and
streams accessible for a diverse range of R users. With straightforward
functionalities, this package allows users to quickly plot and explore
administrative boundaries and rivers and streams in Sri Lanka.
}

\hypertarget{introduction}{%
\subsection{Introduction}\label{introduction}}

The \texttt{ceylon} R package \citep{talagala2023ceylon} conveniently
packages shape files corresponding to the geographic features of Sri
Lanka, enhancing user friendliness for seamless integration and
analysis. This allows for minimising the time spent on data searching
and cleaning efforts. Hence, the package ceylon stands out as a catalyst
for research efficiency. Furthermore, the package supports research
reproducibility, allowing others to independently verify and build upon
the work that utilised the data in this package. The data format easily
integrates with tidyverse packages \citep{wickham2019welcome}, fostering
a smooth workflow.

\hypertarget{data}{%
\subsection{Data}\label{data}}

The data were retrieved from the Humanitarian Data Exchange is a
platform that facilitates the sharing and collaboration of humanitarian
data. The coordinate reference system (CRS) for the data is ``Projected
CRS: SLD99 / Sri Lanka Grid 1999 (CRS code 5234)''. Table \ref{table1}
provides a description of datasets.

\begin{table}[!h]
\begin{center}
\begin{tabular}{|l|l|l|}
\hline
 Dataset & Description & Source \\ \hline
sf\_sl\_0  & country boundary & \url{https://data.humdata.org/} \\ 
province  & province boundaries & \url{https://data.humdata.org/} \\ 
district  & district boundaries  & \url{https://data.humdata.org/} \\ 
sf\_sl\_3   & divisional secretariat boundaries  &  \url{https://data.humdata.org/}\\ 
rivers & Sri Lanka rivers and streams shapefiles & \url{https://data.humdata.org/} \\ \hline
\end{tabular}
\caption{\label{table1}Data Description}
\end{center}
\end{table}

\hypertarget{usage}{%
\subsection{Usage}\label{usage}}

\texttt{ceylon} is available on
\href{https://github.com/thiyangt/ceylon}{GitHub}, and can be installed
and loaded into the R session using:

\begin{verbatim}
install.packages("devtools")
devtools::install_github("thiyangt/ceylon")
\end{verbatim}

\begin{Schunk}
\begin{Sinput}
library(ceylon)
\end{Sinput}
\end{Schunk}

The additional packages required for plotting are as follows:

\begin{Schunk}
\begin{Sinput}
library(ggplot2)
library(sp)
library(sf)
library(viridis)
library(patchwork)
\end{Sinput}
\end{Schunk}

The package ggplot2 \citep{wickham2016ggplot2} is used for data
visualization. The sp \citep{pebesma2005classes} provides tools for
handling spatial data. The sf simple features \citep{pebesma2018simple}
builds upon the strengths of the sp package, introducing efficient
approach to handling spatial data. The viridis \citep{r2023viridis}
package provides a collection of color palettes that are color blind
friendly. The patchwork \citep{pedersen2023patchwork} package is used
for the combination and arrangement of multiple plots. Figure 1 shows
the visualizations of Sri Lanka's administrative borders based on data
available in the \texttt{ceylon} package. The codes to produce Figure 1
is given below.

\begin{Schunk}
\begin{Sinput}
data(sf_sl_0)
a <- ggplot(sf_sl_0) + geom_sf() + ggtitle ("a: Country")
data(province)
b <- ggplot(province) + geom_sf() + ggtitle("b: Province")
data(district)
c <- ggplot(district) + geom_sf() + ggtitle("c: District")
data(sf_sl_3)
d <- ggplot(sf_sl_3) + geom_sf() + ggtitle("d: Divisional Secretariat")
(a|b)/(c|d)
\end{Sinput}
\begin{figure}
\includegraphics{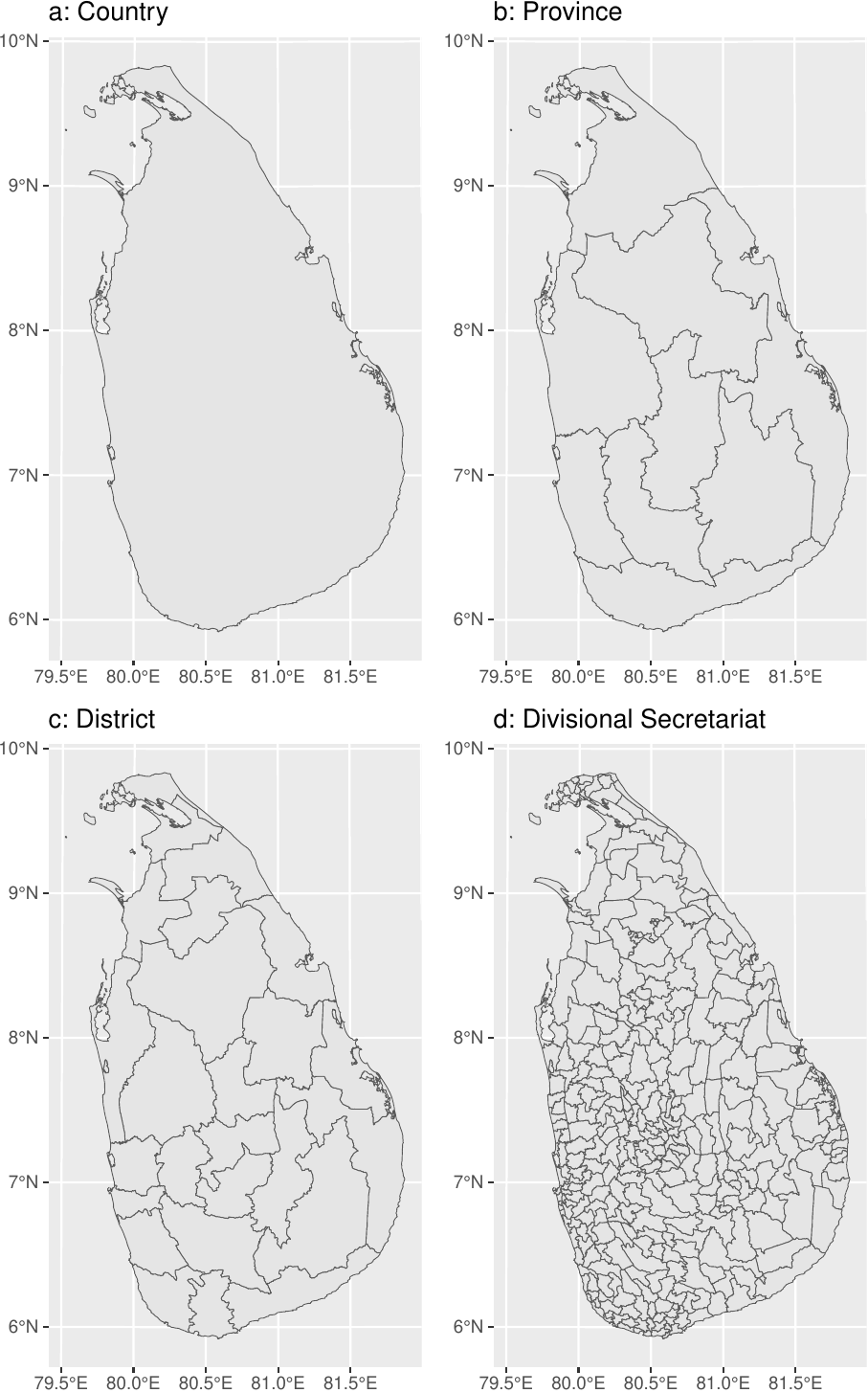} \caption[Maps of differnt administrative divisions in Sri Lanka using data in ceylon package]{Maps of differnt administrative divisions in Sri Lanka using data in ceylon package}\label{fig:unnamed-chunk-3}
\end{figure}
\end{Schunk}

\hypertarget{point-map-adding-a-point-to-the-map}{%
\subsection{Point Map: Adding a point to the
map}\label{point-map-adding-a-point-to-the-map}}

The Global Positioning System (GPS) coordinates of Bandaranaike
International Airport, Sri Lanka is Latitude: 7.1753 Longitude: 79.8835.
The goal is to plot this point along with the province boundaries. The
EPSG:4326 geographic CRS system gives latitude and longitude coordinates
to specify a location on the surface of the earth. Hence, before
plotting, first longitude and latitudes should be converted into sf
object to the same coordinate reference system as the province data set.
For this the sp \citep{pebesma2005classes} and sf
\citep{pebesma2018simple} packages in R were used. In the following code
\texttt{st\_as\_sf} specify the current coordinate reference system for
longitude and latitude. The function \texttt{st\_transform} converts the
current CRS to the target CRS. The target CRS is the CRS associated with
the province, which is defined as \texttt{crs\ =\ st\_crs(province)}
inside the'st\_transform' function.

\begin{Schunk}
\begin{Sinput}
airport <- data.frame(lng = 79.8835, lat = 7.1753)
airport.new <- airport |>
  st_as_sf(coords = c("lng", "lat"), crs = 4326) |>
  st_transform(crs = st_crs(province))
\end{Sinput}
\end{Schunk}

\begin{Schunk}
\begin{Sinput}
point <- ggplot(province) + 
  geom_sf() + geom_sf(data = airport.new, size = 2, col = "darkred") + 
  ggtitle("a: Point")
\end{Sinput}
\end{Schunk}

\hypertarget{line-map-plot-rivers-and-streams-in-sri-lanka}{%
\subsection{Line Map: Plot rivers and streams in Sri
Lanka}\label{line-map-plot-rivers-and-streams-in-sri-lanka}}

This section illustrates an example of using the rivers dataset in the
package ceylon.

\begin{Schunk}
\begin{Sinput}
data("rivers")
line <- ggplot(data = sf_sl_0) +
  geom_sf(fill="#edf8b1", color="#AAAAAA") +
  geom_sf(data=rivers, colour="#253494") +
  labs(title =  "b: Line")
\end{Sinput}
\end{Schunk}

\hypertarget{polygon-map-creating-a-choropleth-map}{%
\subsection{Polygon Map: Creating a choropleth
map}\label{polygon-map-creating-a-choropleth-map}}

A Choropleth map shows different regions coloured according to the
numerical values associated with each individual region.

\begin{Schunk}
\begin{Sinput}
polygon <- ggplot(province) + 
  geom_sf(mapping = aes(fill = population)) + scale_fill_viridis() + 
  ggtitle("c: Polygon")
\end{Sinput}
\end{Schunk}

\begin{Schunk}
\begin{Sinput}
(point|line)
\end{Sinput}
\begin{figure}
\includegraphics{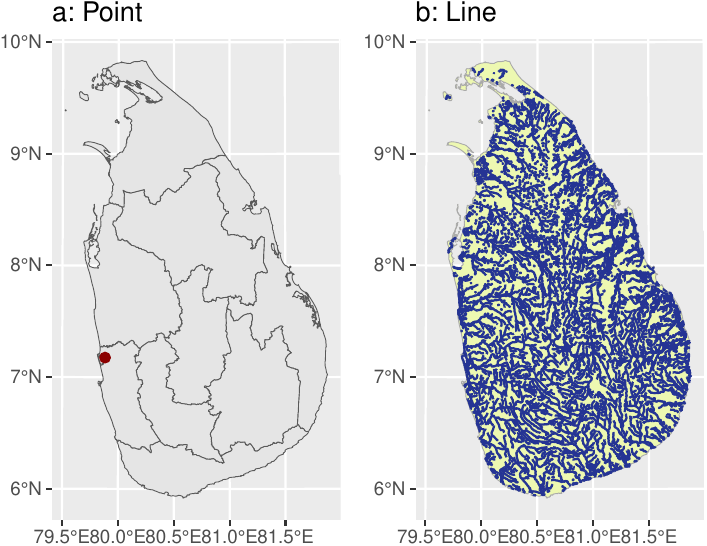} \caption[Point, line, and polygon maps draw using the ceylon package data]{Point, line, and polygon maps draw using the ceylon package data}\label{fig:unnamed-chunk-8}
\end{figure}
\end{Schunk}

\begin{Schunk}
\begin{Sinput}
(polygon)
\end{Sinput}
\begin{figure}
\includegraphics{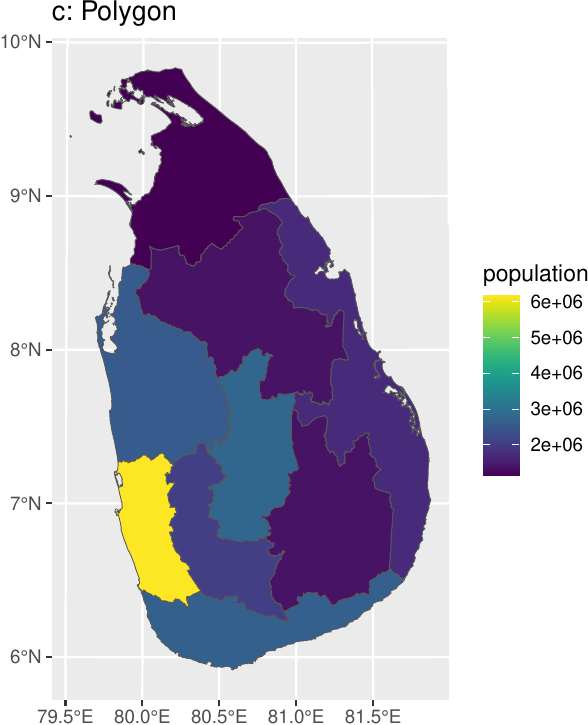} \caption[Polygon maps draw using the ceylon package data]{Polygon maps draw using the ceylon package data}\label{fig:unnamed-chunk-9}
\end{figure}
\end{Schunk}

\newpage

Figure 2 and Figure 3 show the point, line, and polygon maps
created using the data available in the \texttt{ceylon} package.

The above examples illustrate the datasets available in the package,
which are easily integrated with other companion packages that are
widely used in spatial analysis and data visualisation.

\bibliography{RJreferences.bib}

\begin{thebibliography}{7}
\providecommand{\natexlab}[1]{#1}
\providecommand{\url}[1]{\texttt{#1}}
\expandafter\ifx\csname urlstyle\endcsname\relax
  \providecommand{\doi}[1]{doi: #1}\else
  \providecommand{\doi}{doi: \begingroup \urlstyle{rm}\Url}\fi

\bibitem[{Garnier} et~al.(2023){Garnier}, {Simon}, {Ross}, {Noam}, {Rudis}, {Robert}, {Camargo}, Pedr, {Sciain}, {Marc}, {Schere}, and {C\'{e}dric}]{r2023viridis}
{Garnier}, {Simon}, {Ross}, {Noam}, {Rudis}, {Robert}, {Camargo}, A.~Pedr, {Sciain}, {Marc}, {Schere}, and {C\'{e}dric}.
\newblock viridis(lite) - colorblind-friendly color maps for r, 2023.
\newblock URL \url{https://sjmgarnier.github.io/viridis/}.
\newblock viridis package version 0.6.4.

\bibitem[Pebesma(2018)]{pebesma2018simple}
E.~Pebesma.
\newblock Simple features for r: Standardized support for spatial vector data.
\newblock \emph{The R Journal}, 10\penalty0 (1):\penalty0 439--446, 2018.
\newblock \doi{10.32614/rj-2018-009}.
\newblock URL \url{https://doi.org/10.32614/RJ-2018-009}.

\bibitem[Pebesma and Bivand(2005)]{pebesma2005classes}
E.~J. Pebesma and R.~Bivand.
\newblock Classes and methods for spatial data in r.
\newblock \emph{R News}, 5\penalty0 (2):\penalty0 9--13, Nov. 2005.
\newblock URL \url{https://CRAN.R-project.org/doc/Rnews/}.

\bibitem[Pedersen(2023)]{pedersen2023patchwork}
T.~L. Pedersen.
\newblock patchwork: The composer of plots, 2023.
\newblock URL \url{https://CRAN.R-project.org/package=patchwork}.
\newblock R package version 1.1.3.

\bibitem[Talagala(2023)]{talagala2023ceylon}
T.~S. Talagala.
\newblock ceylon: Creating maps of sri lanka administrative regions, rivers and streams, 2023.
\newblock URL \url{https://doi.org/10.5281/zenodo.10432141}.

\bibitem[Wickham(2016)]{wickham2016ggplot2}
H.~Wickham.
\newblock \emph{ggplot2: Elegant Graphics for Data Analysis}.
\newblock Springer-Verlag New York, 2016.
\newblock ISBN 978-3-319-24277-4.
\newblock URL \url{https://ggplot2.tidyverse.org}.

\bibitem[Wickham et~al.(2019)Wickham, Averick, Bryan, Chang, McGowan, Fran\c{c}oi, Grolemun, Haye, Henr, Heste, Kuh, Pederse, Mille, Bach, M\"{u}ll, ~, ~, ~, ~, ~, ~, ~, ~, and ~]{wickham2019welcome}
H.~Wickham, M.~Averick, J.~Bryan, W.~Chang, L.~D. McGowan, R.~Fran\c{c}oi, G.~Grolemun, A.~Haye, L.~Henr, J.~Heste, M.~Kuh, T.~L. Pederse, E.~Mille, S.~M. Bach, K.~M\"{u}ll, J.~O. ~, D.~R. ~, D.~P.~S. ~, V.~S. ~, K.~T. ~, D.~V. ~, C.~W. ~, K.~W. ~, and H.~Y. ~.
\newblock Welcome to the tidyverse.
\newblock \emph{Journal of Open Source Software}, 4\penalty0 (43):\penalty0 1686, 2019.
\newblock \doi{10.21105/joss.01686}.

\end{thebibliography}

\address{%
Thiyanga S. Talagala\\
Department of Statistics\\%
Faculty of Applied Sciences\\ University of Sri Jayewardenepura, Sri
Lanka\\
\url{https://github.com/thiyangt/ceylon}\\%
\textit{ORCiD: \href{https://orcid.org/0000-0002-0656-9789X}{0000-0002-0656-9789X}}\\%
\href{mailto:ttalagala@sjp.ac.lk}{\nolinkurl{ttalagala@sjp.ac.lk}}%
}

\end{article}

\end{document}